# On Bus Ridership and Frequency

Simon J. Berrebi[a,*], Sanskruti Joshi[a], Kari E. Watkins[a]

[a]*Georgia Institute of Technology*
*School of Civil and Environmental Engineering*

**Abstract**

Even before the start of the COVID-19 pandemic, bus ridership in the United States had attained its lowest level since 1973. If transit agencies hope to reverse this trend, they must understand how their service allocation policies affect ridership. This paper is among the first to model ridership trends on a hyper-local level over time. A Poisson fixed-effects model is developed to evaluate the ridership elasticity to frequency on weekdays using passenger count data from Portland, Miami, Minneapolis/St-Paul, and Atlanta between 2012 and 2018. In every agency, ridership is found to be elastic to frequency when observing the variation between individual route-segments at one point in time. In other words, the most frequent routes are already the most productive in terms of passengers per vehicle-trip. When observing the variation within each route-segment over time, however, ridership is inelastic; each additional vehicle-trip is expected to generate less ridership than the average bus already on the route. In three of the four agencies, the elasticity is a decreasing function of prior frequency, meaning that low-frequency routes are the most sensitive to changes in frequency. This paper can help transit agencies anticipate the marginal effect of shifting service throughout the network. As the quality and availability of passenger count data improve, this paper can serve as the methodological basis to explore the dynamics of bus ridership.

## 1. Introduction

Following seven years of consecutive decline, 2019 bus ridership in the United States was the third-lowest year since World War II [1]. Each transit trip lost to private cars contributes to traffic congestion, pollution, and road fatalities. The revenue lost from declining ridership also impedes the ability of transit agencies to provide service, which hurts ridership further in a downward cycle. Transit agencies, therefore, need to understand how this trend can be reversed and at what cost.

The main tools transit agencies have available to influence ridership are service allocation policies. Transit planners are tasked with setting frequencies throughout the network under constrained resources. They must balance ridership with other, sometimes conflicting, objectives including equitable access, connection to places of strategic importance, and reliability. In particular, agencies must decide whether to spread or concentrate service. In order to allocate service in a transparent manner that maximizes total welfare, the effect of frequency on ridership should be quantified. This effect may not be linear. Each vehicle-trip added to a route may not produce as much (or as little) ridership as the current route productivity, measured in passengers per vehicle-trip. The elasticity of ridership with respect to frequency measures the percentage change in ridership resulting from a 1% increase in frequency. When this elasticity is greater than one, adding more service increases the route productivity. Because this elasticity measures the sensitivity of demand to frequency, it varies across routes based on frequency and across stops based on local characteristics.

---

[*]Corresponding author
[1]for a more in-depth examination of recent ridership trends in the United States, see Watkins et al. (2019)

This paper presents a new method to scrub, process, and model ridership data over time and space. Passenger counts are cross-checked with the General Transit Feed Specification (GTFS), a schedule meta-data standard. Passenger counts are then aggregated by route-segments (groups of seven stops on the same route and direction) and combined with data sources on population and jobs. The *change* in ridership is modeled over time through panel regression. Fixed-effects models avoid unobserved heterogeneity and endogeneity biases that plague cross-sectional models by using each individual as its own control over time. Fixed-effects models therefore control for variation **between** individual locations to capture the variation **within** each. In this paper, ridership is modeled using Poisson fixed-effects, which was developed by Hausman et al. (1984) for count data such as passenger boardings and alightings.

The paper proceeds as follows: Section 2 organizes the main studies from the literature by level of aggregation and identifies panel models of hyper-local ridership trends as the gap in research. Section 3 presents the four case studies. Section 4 describes the process of cleaning, aggregating, and combining multiple relevant datasets. In Section 5, the fixed-effects Poisson regression model for ridership data is developed. Section 6 presents the modeling results. Section 7 discusses their implications and identifies future research questions.

**2. Literature Review**

Ridership elasticity to frequency on a local-level and over time remains largely unaddressed in the literature. Table 1 classifies the main studies on transit ridership by level of spatial aggregation (rows) and whether the sample is observed once or at multiple time periods (columns). All the references in the top row evaluate ridership at the transit agency or metropolitan area level. These studies can help compare the impact of aggregated factors across regions, and for studies in the top-right quadrant, across time. Many factors that explain ridership can vary widely between regions and explain why ridership is greater in New York City than in Mobile, AL (Taylor et al., 2009). In a study of 25 regions between 2002 and 2015, Boisjoly et al. (2018) found that the service quantity is among the best predictors of ridership. This relationship, however, seems to have deteriorated in recent years. Following nation-wide cuts in bus service between 2009 and 2012, bus ridership kept falling even as vehicle revenue miles continually increased (Watkins et al., 2019). While these studies shed light on the impact of overall operating expenses on overall ridership, the level of aggregation may drown some of the local dynamics. For example, shifting resources from one route to another can impact overall ridership without even affecting the total service provided. Likewise, the local effects of population, jobs, and demographic factors are diluted in system-level studies.

Table 1: Main studies on transit ridership by level of spatial aggregation (rows) and whether the sample is observed once or at multiple time periods (columns)

|  | Cross-Section | Panel |
|---|---|---|
| System-Level | Taylor et al. (2009); Ingvardson and Nielsen (2018) | Kain and Liu (1999); Kohn et al. (2000); Brown and Thompson (2008); Lane (2010); Chen et al. (2011); Iseki and Ali (2015); Boisjoly et al. (2018); Driscoll et al. (2018); Hall et al. (2018); Graehler Jr et al. (2019); Taylor et al. (2019); Ederer et al. (2019); Watkins et al. (2019); Ko et al. (2019); Lee and Yeh (2019); Woo et al. (2020) |
| Local-Level | Peng et al. (1997); Kimpel et al. (2007); Estupiñán and Rodríguez (2008); Ryan and Frank (2009); Gutiérrez et al. (2011); Pulugurtha and Agurla (2012); Dill et al. (2013); Chakrabarti and Giuliano (2015); Hu et al. (2016); Chakour and Eluru (2016); Sun et al. (2016); Aston et al. (2016); Ma et al. (2018); Mucci and Erhardt (2018); Fasching (2018); Taylor et al. (2019); Wu et al. (2019); Aston et al. (2020b) | Maloney et al. (1964); Kyte et al. (1988); Tang and Thakuriah (2012); Frei and Mahmassani (2013); Kerkman et al. (2015); Brakewood et al. (2015); Berrebi and Watkins (2020); Diab et al. (2020) |

The bottom left quadrant of Table 1 contains studies modeling bus ridership at a local level at a single point in time. These studies capture the local variation **between** individual stops, routes, or route-segments. These models are formulated as if each individual location was similar in every aspect except for the variation captured by explanatory variables. In reality, individual locations are different from each other in ways that cannot be measured and that may affect ridership. Ridership itself is likely to affect frequency directly; when allocating service throughout the network, transit planners strive to maximize ridership. Therefore, estimates for frequency, population, and jobs are likely inflated by the two-way causality. While these estimates inform the relationship between these variables at one point in time, they cannot serve as true elasticities.

The bottom right quadrant contains studies modeling bus ridership on a local level over time. Results from Maloney et al. (1964) and Kyte et al. (1988) still serve as the main reference on ridership elasticity to frequency (Evans et al., 2004; Litman, 2017). Following service changes in Boston, MA, and Portland, OR, these studies computed separate ridership elasticities for each route. With values ranging from zero to 3.77 (Kyte et al., 1988), route-specific elasticities are not entirely meaningful.

Fixed-effects models harness the combined explanatory powers of all individual locations in the panel to explain the change in ridership over time. Tang and Thakuriah (2012) and Brakewood et al. (2015) use fixed-effects models at the route level to evaluate the impact of real-time passenger information on ridership. More recently Diab et al. (2020) modeled the impact of service attributes on ridership between 2012 and 2017 in Montreal, QC. They find a positive correlation with vehicle speed, number of stops, job accessibility, and several measures of service quantity, including frequency, a dummy variable for routes with headways less than 10 minutes, and a dummy variable for the express network. Frei and Mahmassani (2013) and Kerkman et al. (2015) model ridership change at the stop-level over a one year period. These studies apply a log-transform to reduce the asymmetry in the ridership distribution. However, The magnitude of logged-variation is typically far greater at low-ridership stops, which are correlated with explanatory variables by design. In addition, zero-values in the original data must be either truncated or a constant must be added to each observation, which can affect the distribution of variation and produce further heteroskedasticity (King, 1988; Manning and Mullahy, 2001; Silva and Tenreyro, 2006).

Modeling ridership change at a disaggregated level is complicated by the missing data, endogeneity, and multiple levels of interaction. The models are highly sensitive to even slight misspecification. This paper presents a framework to analyze the causes of ridership change on a disaggregated level. A novel methodology to process, combine, and model passenger count data over multiple periods is developed. The Poisson Fixed-Effects model for ridership analysis is introduced. This model is specifically designed to reveal the sensitivity of ridership to frequency on a highly disaggregated spatial and temporal scale without truncating or transforming the data. Retrospectively, this analysis can be used to control for frequency, and identify underlying trends based on a host of other factors as in Berrebi and Watkins (2020). More broadly, this study opens the door to a wide range of research topics on the sensitivity of transit ridership.

### 3. Case Studies

To evaluate the relationship between transit ridership and frequency, four transit agencies were selected due to the high quality of their APC data:

- Tri-County Metropolitan Transportation District of Oregon (TriMet) in Portland OR
- Miami-Dade Transit in Miami, FL
- Metro Transit in Minneapolis/St-Paul, MN
- Metropolitan Atlanta Rapid Transit Authority (MARTA) in Atlanta, GA

The study team initially approached transit agencies asking for historical Automatic Passenger Count (APC) data going back as far as possible by markup, which is the period of schedule planning (also known as pick, shakeup, or signup). Transit agencies typically have three markups per year, spring, summer, and fall/winter. Since agencies started implementing APC technology at different times, the range of available data varies. Table 2 shows the first and last markup for each agency. Every markup in between was used in this analysis.

Table 2: First and last markup selected for each agency

|  | TriMet | Miami-Dade | Metro Transit | MARTA |
|---|---|---|---|---|
| First markup | spring 2012 | fall/winter 2013 | fall/winter 2012 | summer 2014 |
| Last markup | spring 2017 | fall/winter 2017 | fall/winter 2017 | summer 2018 |

Table 3 shows the characteristics of the transit agencies and their metropolitan areas. Data in the first two rows, 2018 bus ridership and hours of directly operated service [2], come from the National Transit Database (Federal Transit Administration, 2019). The last two rows of Table 3 show Metropolitan Statistical Area (MSA) population and percent of population living in dense Census Tracts according to the 2016 American Community Survey (US Census Bureau, 2016). Percent living in density is calculated as the share of metropolitan area population living in Census Tracts with more than three housing units per gross acre, which corresponds to the density threshold that can financially justify running transit service as calculated by Pushkarev and Zupan (1977).

Table 3: Ridership and service provided by agency

|  | TriMet | Miami-Dade | Metro Transit | MARTA |
|---|---|---|---|---|
| Yearly Unlinked Passenger Trips (000's) | 56,727 | 49,716 | 54,910 | 49,788 |
| Yearly Vehicle Revenue Hours (000's) | 1,988 | 1,961 | 2,050 | 2,249 |
| MSA population (000's) | 2,425 | 6,066 | 3,551 | 5,790 |
| % living in transit-supportive density | 43.1 | 58.7 | 22.9 | 10.8 |

The transit agencies in this study are similar in size but they vary widely in other aspects. Ridership and revenue hours in each of the four agencies are all within 15% of each other. Dividing passenger trips by revenue hours gives the productivity. TriMet is the most productive agency with 28.5 passengers per revenue hour and MARTA is the least productive with 22.1 passengers per revenue hour. While TriMet serves more passenger trips than Miami-Dade, the Miami region has 2.5 times more population than the Portland area. The Miami region is the densest, followed by Portland, Minneapolis/St-Paul, and Atlanta, where only 11% of the population lives at transit-supportive densities. The case-studies, therefore, represent a wide cross-section of mid-sized transit agencies, forming a basis of comparison that can be useful to their peers.

---

[2] Contracted service is not considered in this study because the passenger count data differs from directly operated service in accuracy, format, and availability.

## 4. Data Processing

Transit agencies in this study provided the research team with ridership data at the stop-route-direction-trip level. Passenger count data were then scrubbed, aggregated, and combined with other data-sets through a process illustrated in Figure 1. Each step of this process is described in this section. The outcome is a dataset that serves as input to our models. For a detailed description, see Joshi (2019). Note that this same process could be applied to Automated Fare Collection (AFC) data.

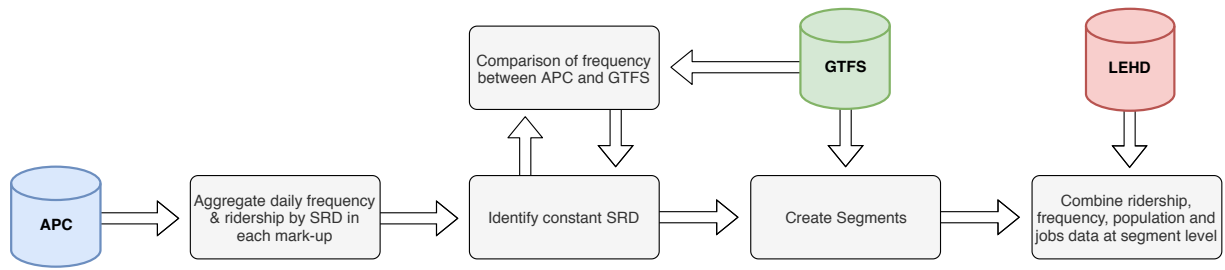

Figure 1: Flow chart describing the data scrubbing and aggregation process

### 4.1. Aggregate Daily Frequency and Ridership

To provide a basis of comparison between all possible combinations of stop-route-direction (SRD), total daily frequency and ridership were aggregated by day. Total daily frequency was obtained by counting the number of trips. Aggregated ridership was obtained by averaging the sum of boardings and alightings across all weekday trips in the markup. Summing both boardings and alightings is necessary to avoid the asymmetry problem: some stops, typically located near the end of the line, are only ever used to alight buses. For example, 18% of stops in Minneapolis/St-Paul have zero recorded boardings in fall 2012. In addition, all weekdays are pooled together, their averages are based on sample sizes five times greater than Saturdays and Sundays. In order to consider the most comprehensive dataset possible, we chose to focus on weekday ridership.

### 4.2. Identify Constant Stop-Route-Directions

In order to understand the relationship between the *change* in ridership and the *change* in frequency, only stop-route-directions that remained constant over the entire study period were considered. Any stop that was either added or removed between the first and the last markup was disregarded.

### 4.3. Comparison Between APC and GTFS

Passenger counts cannot be relied upon as the sole source of frequency data because trips are commonly missing[3]. Transit agencies usually purchase vehicles already equipped with APC. Older vehicles, however, do not produce passenger counts. Even on new vehicles, APC units can go out of service for weeks or months before getting repaired. The same bus is often scheduled to operate a trip every day of the markup. If that vehicle is not equipped with a functioning APC, then the trip will never be surveyed. Missing trips can introduce a bias. As older vehicles get rolled out of the fleet, APC coverage increases over the years, which could be interpreted as an increase in frequency.

---

[3]Fortunately, the passenger counts themselves are reliable. The four transit agencies have systems in place to ensure that trips are not under-counted or over-counted. A study in TriMet from the late 1980s had already found the APC data to be consistent with manual counts (Strathman and Hopper, 1989).

To avoid this issue, the number of observed daily trips in APC was compared with the number of scheduled trips for each stop-route direction. Historical schedule data published by transit agencies in GTFS format were obtained from third-party websites [4]. The number of daily trips in APC and GTFS was then compared. All segments with more than one missing trip in the first markup were filtered out of the analysis entirely. Route-segments with more than one missing trip in subsequent years are removed from the analysis only for the offending years. For an in-depth analysis comparing APC and GTFS, see Berrebi et al. (2021). The sensitivity of the results to missing trips is analyzed in Appendix A.

*4.4. Route-Segments*

While the main objective of this paper is to understand the relationship between frequency and ridership, population and jobs also affect transit ridership on a local level. To extract meaningful results, these variables should be measured on the scale of their variation. While the service coverage area (i.e. accessible walking distance) for a bus stop is defined by the Transit Capacity and Quality of Service with a $\frac{1}{4}$ mi radius (Kittelson Associates, 2013, §5-10), typical stop spacing in urban areas is only $\frac{1}{8}$ mi according to the TCRP Report 19 - Location and Design of Bus Stops (Fitzpatrick et al., 1996). Hyper-local variations in ridership are more likely to be explained by walkability, which is determined by connectivity, land-use patterns, quality of path, and context, on which no data is available (Southworth, 2005). Therefore, what accounts for differences in ridership between adjacent stops on the same route-direction is unlikely to be captured by our explanatory variables. Modeling ridership at the stop-route-direction level would reduce explanatory power. Furthermore, the passenger's choice to use one stop over the next one introduces serial correlation, which may affect estimated variances.

While minimizing the overlap between adjacent segments, it is also important to maintain locally relevant units of spatial analysis. Segments must be short enough for the population and jobs collected within a 1/4 mi radius to describe the potential demand on a local level. To address these issues, we define route-segments, clusters of seven adjacent stops on the same route-direction, as the spatial unit of analysis.

Route-segments are created using the stop sequence (i.e. order on the route) of constant stop-route-directions, which was obtained from GTFS. Since a stop can have different sequences on the same route-direction across different trips, the stops, stop_times, trips, and routes tables were joined and the sequence index common to the most trips was recorded. Ultimately, each constant stop-route-direction combination was present in one and only one route-segment. Spatial coordinates for each stop were also obtained from the GTFS stops table. Each route-direction was then divided into segments of seven stops in sequential order. The last remaining stops at the end of each route were merged with the upstream segment. Ridership and frequency were then averaged by segment across stops.

---

[4] https://transitfeeds.com/
http://www.gtfs-data-exchange.com/

*4.5. Population and Jobs*

Population and job data were obtained from Longitudinal Employer-Household Dynamics (LEHD). The LEHD are data products compiled by states using Unemployment Insurance earnings and published by the US Census Bureau. The number of jobs is provided at the Census Block level, by year. These data, however, are only available between 2011 and 2015. We therefore linearly extrapolated the LEHD data to match the APC time-frame. The underlying assumption is that population and job trends between 2011 and 2015 continued their course until 2018 in Atlanta and until 2017 everywhere else. In other words, a Census Block that gained five residents per year until 2015 was assumed to keep growing at the same rate.

The LEHD data were first cleaned and prepared for import into ArcMap. Block Group shapefiles were joined using a common GEOID field. A dissolved $\frac{1}{4}$ mi buffer was applied to the stops based on the common route-segment field. This buffer was overlaid with LEHD data using a pairwise intersection tool, which compares the input features of overlapping layers. Features common to both input layers were sent to the output feature class. The output mirrors the geometric intersections of the two layers while considering which layers they are derived from. Therefore, the overlapping service coverage areas of bus stops within the same segment were only counted once. Note, however, that service areas falling in the overlap of multiple segments were assigned to each of the segments. Under the assumption that people and jobs are uniformly distributed within their geographic unit, LEHD data were weighted by the proportion of each Census Block in the segment buffer.

Figure 2 shows a map in Portland, OR, illustrating LEHD workplace locations by route-segment buffers, which are differentiated by color. Overlapping segment buffers hide each other and are therefore not visible on the map. The length of each buffer varies considerably depending on the stop density. Block Group boundaries are shown as thick dashed lines and Census Block boundaries are shown as light gray lines. The small gaps in segment buffers represent the Census Blocks without any job locations.

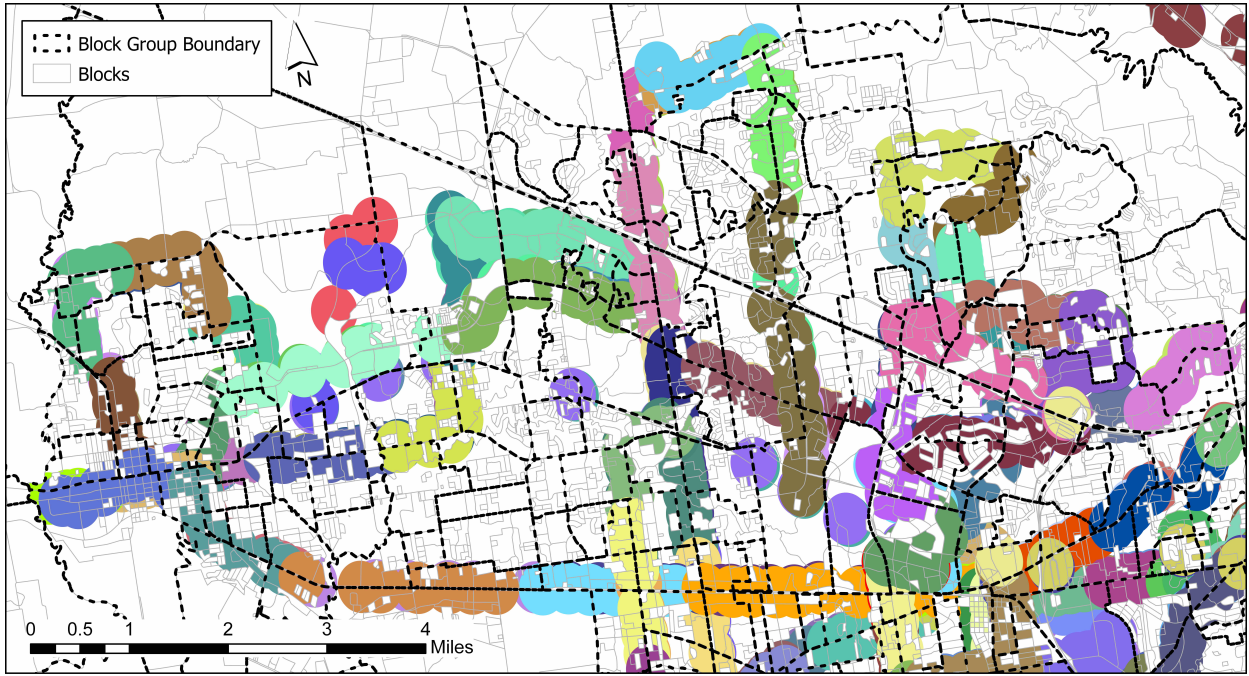

Figure 2: Map illustrating LEHD workplace locations by route-segment buffers along with Census Blocks and Block Groups lines

## 5. Model

### 5.1. Regression Structure

Ridership data is count data[5]. These data violate the Ordinary Least Square (OLS) assumption of normally distributed errors. While the Normal distribution is defined from $-\infty$ to $\infty$ and has constant variance, count data are necessarily non-negative and their variance grows with their mean. Skewed data, however, can be transformed by taking the natural logarithm (log) to fit OLS as long as they have constant variance-to-mean ratio (Nelder and Wedderburn, 1972). The log-linear model for ridership is given in Equation (1). Variables are defined in Table 4.

$$\log(\text{Rid}) = \beta_0 + \beta_1 \log(\text{Freq}) + \beta_2 \log(\text{Pop+Job}) + \epsilon \qquad (1)$$

In Equation (1), Freq and Pop + Job are logged. Logged variables are measures of exposure. Their coefficients, $\beta_1$ and $\beta_2$ are elasticities. They explain the proportional change in ridership resulting from the proportional change in explanatory variables. Here the number of residential homes and job locations within $\frac{1}{4}$ mi of bus stops are added to form a potential trip generation variable.

---

[5]The response variable in our model, sum of average weekday passenger boardings and alightings, is not discrete. This is because transit agencies typically store stop-level ridership data as averages across entire markups (usually three or four months). Some agencies do not record the number of vehicle-trips these averages are based on. Otherwise, we could have multiplied average ridership by the number of vehicle-trips and treated the latter variable as the offset. In any case, this does not affect the consistency of our model (Silva and Tenreyro, 2006).

Table 4: Summary of variable definitions

| Variable | Definition |
|---|---|
| Rid | Total weekday passenger boardings and alightings |
| Freq | Total weekday vehicle-trips |
| Pop+Job | Total population and jobs within $\frac{1}{4}$ mi of segment |
| $t$ | Year $\in (0, ..., T)$ |
| $i$ | Route-segment $\in (0, ..., n)$ |

Taking the exponent of both side sides of (1) yields the more intuitive formulation (2), in which, explanatory variables have a multiplicative relationship. The model can be summarized as follows: when a bus passes by a route-segment, each residential home and job has the potential to generate a trip. This formulation is consistent with the literature: the effect of frequency on ridership is dependent on population and jobs (Badoe and Miller, 2000). A route running through areas with low population and few jobs is unlikely to attract ridership, even if it has high frequency. Conversely, a route-segment passing through a high concentration of population and jobs still needs frequency to attain high ridership.

$$\text{Rid} = e^{\beta_0} * \text{Freq}^{\beta_1} * (\text{Pop+Job})^{\beta_2} * e^{\epsilon} \qquad (2)$$

While the log-linear OLS regression of Equation (1) gives an estimate of E(log(Rid)), we are primarily interested in log(E(Rid)). A model that estimates log(E(Rid)) directly and deals with zero-values adequately is the Poisson regression. In the Poisson regression, the expected ridership is estimated directly as per Equation (3) through Maximum Likelihood Estimation (MLE). The vectors $\boldsymbol{x}$ and $\boldsymbol{\beta}$ represent the explanatory variables and their coefficients with the exception of the intercept, $\beta_0$.

$$E(\text{Rid}|\boldsymbol{x}) = e^{\beta_0 + \boldsymbol{\beta}\boldsymbol{x}'} \qquad (3)$$

Given a (possibly non-linear) model specification, MLEs estimate the distribution parameters that maximize the probability of obtaining the observed data. In our model, the Poisson regression is computed with a robust covariance matrix (also known as quasi-Poisson regression) to relax the assumption that the variance must equal the mean. The quasi-Poisson model returns the same coefficients as the regular Poisson regression, but standard errors are better calibrated for potential over-dispersion (Wooldridge, 2002, §19.2). Note that our standard errors may still be inflated due to spatial correlation between route-segments. This effect, however, is limited by the clustering of adjacent stops, which leaves little overlap between segments on the same route and direction.

Equation (4) shows the derivative of E(Rid|$\boldsymbol{x}$) with respect to frequency. This derivative is the expected ridership gained from adding a single vehicle-trip to the route-segment, henceforth referred to as "marginal productivity". In the original model, $\beta_1$ is the elasticity of ridership to frequency.

$$\frac{\delta \text{E}(\text{Rid}|\boldsymbol{x})}{\delta \text{Freq}} = \beta_1 * \text{Freq}^{\beta_1 - 1} * e^{\beta_0} * (\text{Pop} + \text{Job})^{\beta_2} \qquad (4)$$

In the derivative of Equation (4), we see that the marginal productivity is proportional to Freq$^{\beta_1 - 1}$. Therefore, when $\beta_1 > 1$ ($\beta_1 < 1$), then the marginal productivity is a monotone increasing (decreasing) function of Freq, i.e. each additional trip yields increasing (decreasing) ridership returns. Identifying the sign of ($\beta_1 - 1$) can help inform the trade-off between service allocation policies.

*5.2. Fixed Effects*

Let us consider the observed ridership, Rid$_{it}$, on route-segment $i$ and year $t$, where $i \in \{1, ..., n\}$ and $t \in \{1, ..., T\}$. The standard formulation for the fixed-effects model is shown in Equation (5). The linear equation $\boldsymbol{\beta} \boldsymbol{x}'_{it}$ is the effect of explanatory variables $\boldsymbol{x}_{it}$. The terms $\alpha_i$ represents the individual specific effects, $\mu$ is a linear time trend, and $\epsilon_{it}$ is the error.

$$\log(\text{Rid}_{it}) = \boldsymbol{\beta} \boldsymbol{x}'_{it} + \alpha_i + \mu t + \epsilon_{it} \qquad (5)$$

The $\alpha_i$ term allows each individual route-segment to have its own intercept, while the $\beta$ terms force the relationship between $\boldsymbol{x}'_{it}$ and Rid$_{it}$ to be the same for all. The term $\mu$ represents the linear effect of time, which is not explained by other variables. It takes into account conjunctural phenomena that may be affecting all route-segments equally.

Similarly to the cross-sectional case in the last section, panel ridership data can be modeled with a Poisson fixed effects model[6]. As in Equation (3), the expected ridership is modeled as follows:

---

[6]As in the cross-sectional case, the fixed-effects Poisson model assumes that the variance, $V(\text{Rid}_{it}|\boldsymbol{x}_i)$, is equal to its mean, $E(\text{Rid}_{it}|\boldsymbol{x}_i)$. To relax this assumption, Hausman et al. (1984) introduced the fixed-effects Negative-Binomial model, which estimates a dispersion parameter. Like Hausman et al. (1984)'s Poisson fixed-effects, their Negative Binomial fixed-effects model is conditioned on the sum of outcomes. This model, however, was shown to be misspecified by Allison and Waterman (2002) and later in more detail by Guimaraes (2008). The Negative-Binomial fixed-effects can also be estimated unconditionally (i.e. by fitting dummy variables, called incidental parameters, for each individual except for one). However, there is, to the authors' knowledge, no proof that it does not (or does) cause an incidental parameter problem. In the unconditional Negative Binomial fixed-effects model, the incidental parameter is not divided out. In cases, such as ours, where T is fixed and N is large, this parameter can be inconsistent and bias the $\beta$ estimates (Hsiao, 2003, §7.3). Fortunately, Wooldridge (1999) showed that the Poisson fixed-effects estimates are robust to overdispersion.

$$E[\text{Rid}_{it}|\boldsymbol{x}_{it}] = e^{\boldsymbol{\beta}\boldsymbol{x}'_{it}+\alpha_i+\mu t} \tag{6}$$

The $\beta$ parameters are estimated conditionally on the sum of outcomes at the individual level over the years (Hausman et al., 1984). The resulting conditional likelihood is proportional to the right-hand side of Equation (7) as per (Cameron and Trivedi, 1998, §9.3). Estimates are obtained by setting the partial derivative of the log-likelihood with respect to beta to zero as in Equation (8).

$$\mathcal{L}_c(\boldsymbol{\beta}) \propto \sum_{i=1}^{n}\sum_{t=1}^{T} \text{Rid}_{it} \log\left(\frac{e^{\boldsymbol{\beta}\boldsymbol{x}'_{it}}}{\sum_{s=1}^{T} e^{\boldsymbol{\beta}\boldsymbol{x}'_{is}}}\right) \tag{7}$$

$$\sum_{i=1}^{n}\sum_{t=1}^{T} \boldsymbol{x}_{it}\left(\text{Rid}_{it} - \sum_{s=1}^{T}(\text{Rid}_{is})\frac{e^{\boldsymbol{\beta}\boldsymbol{x}'_{it}}}{\sum_{s=1}^{T} e^{\boldsymbol{\beta}\boldsymbol{x}'_{is}}}\right) = 0 \tag{8}$$

Notice that any constant added to the observed explanatory variables, $\boldsymbol{x}_{it}$, gets canceled out in both equations. Therefore, only the variation taking place over time **within** each individual route-segment is modeled. Any heterogeneity and endogeneity disappear as long as it does not vary over time. Because spatial demand and the resulting allocation of service change at the pace of land-use, we are not concerned about these effects over a five to six-year panel.

### 5.3. Isolating the Effects of Frequency Change on Ridership

Time-invariant explanatory variables cannot be considered in fixed-effects models. However, some of the research questions raised in the introduction compel us to evaluate the relationship between prior frequency (frequency at the beginning of the panel, when $t = t_0$) and the change in transit ridership. Time-invariant variables, in this case prior frequency, can be interacted with time-varying variables, in the case frequency at time $t$. The Poisson fixed-effects model considered in this paper is shown in Equation (9). Noticing that logs cancel their exponents, we obtain the more intuitive formulation of Equation (10).

$$E[\text{Rid}_{it}|\boldsymbol{x}_{it}] = e^{\left(\begin{array}{c}\beta_1 \log(\text{Freq}_{it}) + \beta_2\left[\log(\text{Freq}_{it_0})*\log(\text{Freq}_{it})\right] \\ +\beta_3 \log(\text{Pop}_{it}+\text{Job}_{it})+\alpha_i+\mu t\end{array}\right)} \tag{9}$$

$$= \text{Freq}_{it}^{\beta_1+\beta_2\log(\text{Freq}_{it_0})} * (\text{Pop}_{it}+\text{Job}_{it})^{\beta_3} * e^{\alpha_i} * e^{\mu t} \tag{10}$$

Frequency ($\text{Freq}_{it}$), Population ($\text{Pop}_{it}$), Jobs ($\text{Job}_{it}$), and time ($t$) are time-varying, while prior frequency is time-invariant ($\text{Freq}_{it_0}$). The interpretation of $\beta_1$ is the elasticity of ridership to frequency, as in the cross-section model. However, sensitivity is measured **within** individual route-segments instead of **between**. The second term is the interaction of the frequency at $t = t_0$ with the time-varying frequency. The coefficient $\beta_2$ tells us whether ridership is more elastic to frequency on frequent routes (positive) or on infrequent routes (negative). Finally, the fourth term, $\beta_3$, gives the **within** elasticity to change in population and jobs.

Equation (10) can be used to predict how changing frequency on a route-segment would affect ridership. However, this relationship is dependent on the individual-specific error, $\alpha_i$. This term is divided out in Equation (11).

$$\frac{E[\text{Rid}_{iT}|\boldsymbol{x}_{iT}]}{E[\text{Rid}_{it_0}|\boldsymbol{x}_{it_0}]} = \left(\frac{\text{Freq}_{iT}}{\text{Freq}_{it_0}}\right)^{\beta_1+\beta_2\log(\text{Freq}_{it_0})} * \left(\frac{\text{Pop}_{iT}+\text{Job}_{iT}}{\text{Pop}_{it_0}+\text{Job}_{it_0}}\right)^{\beta_3} * e^{\mu T} \quad (11)$$

As in (4), taking the derivative of $E[\text{Rid}_{iT}|\boldsymbol{x}_{iT}]$ with respect to $\text{Freq}_{iT}$ yields the marginal productivity. Unlike the elasticity, which measures the percentage change in ridership as a function of percentage change in frequency, the marginal productivity measures the absolute change as a function of the absolute change. When $\text{Freq}_{iT} = \text{Freq}_{it_0}$, the marginal productivity is:

$$\frac{\delta E[\text{Rid}_{iT}|\boldsymbol{x}_{iT}]}{\delta \text{Freq}_{iT}} = E[\text{Rid}_{it_0}|\boldsymbol{x}_{it_0}] * \frac{\beta_1+\beta_2\log(\text{Freq}_{it_0})}{\text{Freq}_{it_0}} * \left(\frac{\text{Pop}_{iT}+\text{Job}_{iT}}{\text{Pop}_{it_0}+\text{Job}_{it_0}}\right)^{\beta_3} * e^{\mu T} \quad (12)$$

Equation (12) enables schedule planners to anticipate the effects of service allocation policies just by knowing the prior ridership and frequency. While the static $E[\text{Rid}_{it_0}|\boldsymbol{x}_{it_0}]$ is modeled in Subsection 6.1, using observed values of prior ridership, $\text{Rid}_{it_0}$, instead allows to capture the unobserved heterogeneity. This formulation enables planners to assess how much ridership will be lost by removing one vehicle from a specific route and how much will be gained by adding it to a different route.

## 6. Results

This section presents the results from the Poisson cross-section and fixed-effects models. These models were run in the software program R using the stats and pglm packages, respectively (R Core Team, 2017; Croissant, 2017).

### 6.1. Cross-Section

Table 5 shows the results of the cross-sectional model in the first markup (2012/2013/2014) presented in Equation (2). McFadden's pseudo-$R^2$ is shown for each agency at the bottom of Table 5. In Portland, Miami, and Minneapolis/St-Paul, pseudo-$R^2$ values close to one indicate a good fit; less so in Atlanta. The first row, log(Freq) is the elasticity of ridership to frequency. For all four agencies, elasticity is significantly greater than one, and hence elastic. In other words, for two route-segments with the same population and jobs, the one with the greater frequency is likely to have more passengers per trip. The second row shows the sensitivity of ridership to the total population and jobs. In all four agencies, the parameter estimate is significantly below one. For the same level of frequency, a segment surrounded by more development than another is expected to have less ridership per capita when controlling for frequency. These results are consistent with a meta-study from Aston et al. (2020a). Based on a review of 90 articles, the authors had concluded that ridership was inelastic to population and job density. .

Table 5: Cross Section Models

|  | Response Variable: | Rid | | |
| --- | --- | --- | --- | --- |
|  | Portland | Miami | Minneapolis/St-Paul | Atlanta |
| log(Freq) | 1.36 (0.03)*** | 1.21 (0.03)*** | 1.50 (0.04)*** | 1.33 (0.08)*** |
| log(Pop + Job) | 0.52 (0.02)*** | 0.53 (0.02)*** | 0.52 (0.02)*** | 0.33 (0.04)*** |
| Intercept | −4.87 (0.19)*** | −4.45 (0.17)*** | −5.86 (0.19)*** | −3.57 (0.40)*** |
| Pseudo $R^2$ | 0.82 | 0.72 | 0.81 | 0.39 |
| Deviance | 9447.14 | 13138.63 | 10732.59 | 14363.80 |
| Num. obs. | 874 | 1165 | 959 | 718 |

***$p < 0.001$; **$p < 0.01$; *$p < 0.05$; ·$p < 0.1$

In order to verify the model results of Table 5, we need to represent the relationship between ridership and frequency in the first markup (2012/2013/2014) graphically. Figure 3 shows route-level productivity in weekday passenger boardings per trip against frequency in daily trips. Productivity allows us to compare the ridership contribution of each vehicle-trip on routes with different frequencies. The size of each dot corresponds to the number of stops on the route. As one would expect, routes with more stops connect to more places, and therefore yield more passengers per trip. Express routes in Minneapolis/St. Paul are excluded because their stop densities are so much lower than the rest. A trend line is shown in red. The gray band represents its standard error.

Figure 3 provides a different lens to observe the same phenomenon identified in the cross-sectional regression model. There is a clear positive relationship between productivity (in passenger boardings per trip) and frequency (in number of daily trips) at one point in time. This trend is strongest in Portland and Miami and weakest in Minneapolis. In all four agencies, frequent routes carry more passengers per trip than lower frequency routes. These results are consistent with Mohring (1972) who suggested that frequency should be set as the square root of ridership per mile.

Since these routes concentrate the most service, slight changes in productivity can have a disproportionate impact on overall ridership. If the high productivity of these frequent routes could be maintained, increasing service would greatly benefit overall ridership. On the other hand, transit agencies may want to add service for the specific purpose of reducing crowding. It is indeed important to consider other criteria than productivity only in decisions for frequency increases, such as (maximum) crowding levels.

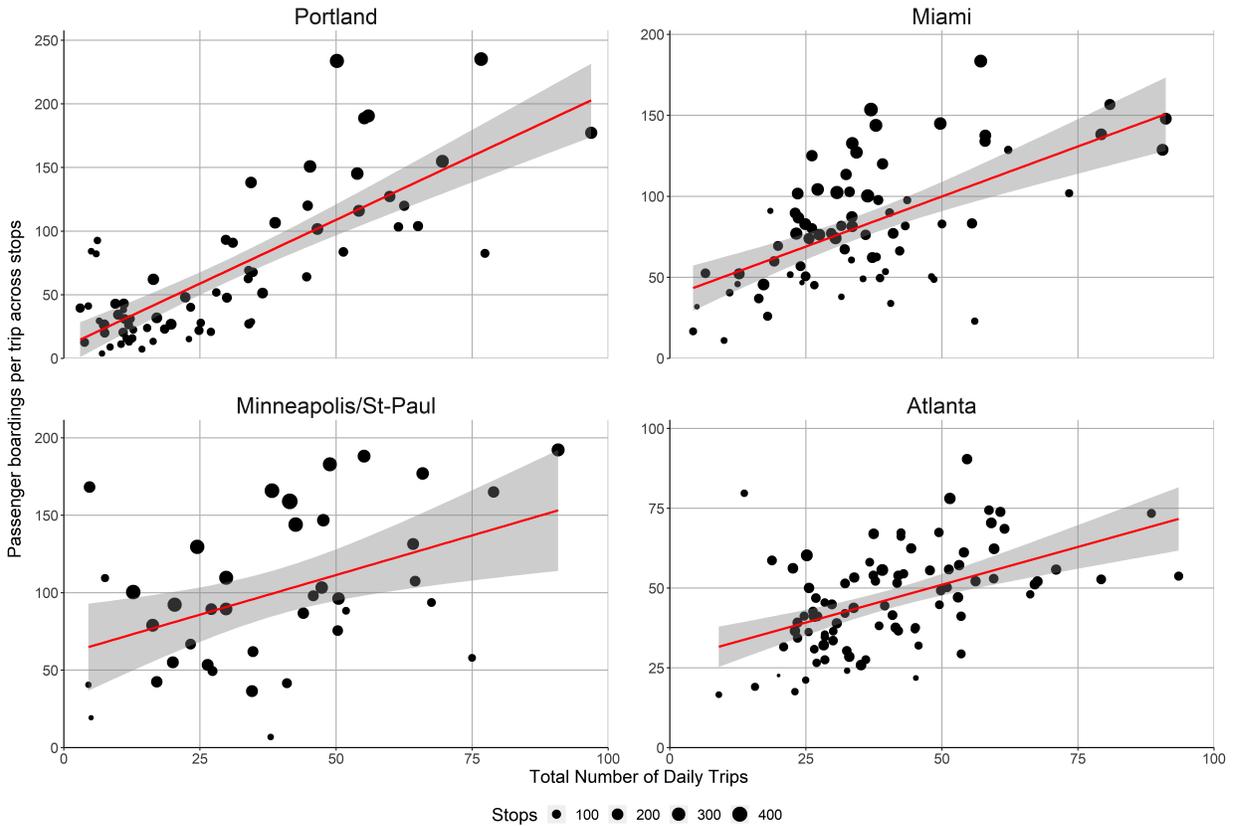

Figure 3: Route-level productivity in passenger boardings per stop per trip as a function of weekday frequency over time in four metropolitan areas in the first markup (2012/2013/2014).

Both the cross-sectional model and the graphical representation show that, when comparing the variation **between** route-segments or routes, productivity is positively correlated with frequency. As discussed in the literature review, this does not necessarily mean that increasing frequency on a route-segment will produce increasing returns. High-frequency segments may be more productive due to unobserved heterogeneity or endogeneity. The effect of frequency change on ridership can, therefore, only be tested by comparing the variation **within** each route-segment *over time*. The next subsection does just that.

6.2. *Fixed Effects*

Table 6 shows the results of the Poisson fixed-effects model. For each agency, the elasticity of ridership to frequency and to population and jobs is presented. In this model, the interaction between frequency and prior frequency is omitted to enable a comparison with the cross-sectional results. The log-likelihood, total number of observations, individual segments, and time periods are shown at the bottom of the table. Unfortunately, there is no equivalent to the pseudo-$R^2$ for the fixed-effects Poisson model. However, the good fit of the cross-sectional models indicates that the fixed-effects model is also well specified. Finally, note that the linear time trend is significantly negative for all agencies. This indicates that, even when controlling for frequency, population, and jobs, weekday ridership is still declining over time.

The elasticity of ridership to frequency is far weaker in the fixed-effects model than in the cross-section. While the **between** elasticity in the previous model ranged from 1.21 to 1.50, the **within** elasticity shown in Table 6 ranges from 0.66 to 0.78. For all agencies studied, ridership is inelastic to frequency. In other words, each vehicle-trip added to a route-segment generates diminishing productivity returns.

Table 6: Fixed Effects Model Without Prior Frequency

|  | Response Variable: | $\text{Rid}_{it}$ | | |
| --- | --- | --- | --- | --- |
|  | Portland | Miami | Minneapolis/St-Paul | Atlanta |
| $\log(\text{Freq}_{it})$ | 0.72 (0.02)*** | 0.78 (0.03)*** | 0.77 (0.02)*** | 0.66 (0.02)*** |
| $\log(\text{Pop}_{it} + \text{Job}_{it})$ | −0.01 (0.02) | 0.04 (0.02)· | 0.05 (0.02)* | 0.10 (0.03)*** |
| t | −0.02 (0.00)*** | −0.05 (0.00)*** | −0.02 (0.00)*** | −0.07 (0.00)*** |
| Log-Likelihood | −35170.06 | −15141.52 | −34303.11 | −11420.78 |
| Num. obs. | 17640 | 10575 | 18256 | 7283 |
| n | 874 | 1165 | 959 | 718 |
| T | 21 | 10 | 21 | 12 |

***$p < 0.001$; **$p < 0.01$; *$p < 0.05$; ·$p < 0.1$

In order to represent the **within** elasticity of ridership to frequency graphically, we must add the time dimension to our scatter plots. Figure 4 shows route-level productivity as a function of frequency in the first markup (2012/2013/2014) in red and the last markup (2017/2018) in blue. The slopes of the arrows linking the first markup and the last markup data points are marginal productivity. There is clearly a trend pulling productivity down in all agencies, which is not dependent on frequency change. However, if frequency had no impact on productivity (i.e. elasticity = 1), then all routes would lose the same relative productivity. In all agencies, routes in which frequency increased seemed to experience more relative decline in productivity than expected and *vice versa*. This is particularly true in Portland, where long arrows pointing towards the bottom-right contrast with short arrows spread in every direction.

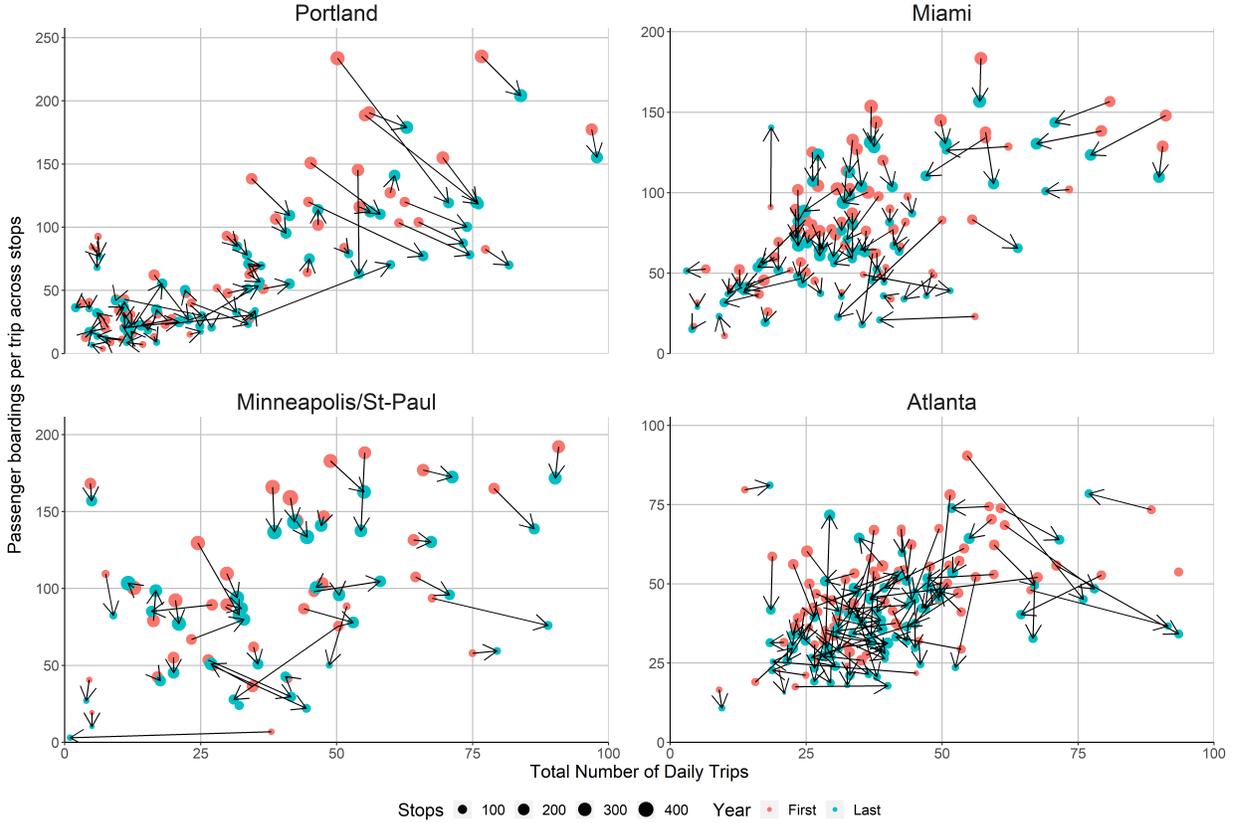

Figure 4: Route level productivity in passenger boardings per stop per trip as a function of weekday frequency over time in four metropolitan areas.

The trends in Figure 4 also show the connection between frequency change and prior frequency. For example, in Portland, frequent routes gained frequency, perhaps in an attempt to combat overcrowding, while in Miami, frequent routes lost the most frequency, perhaps in an attempt to maintain coverage. Adding the interaction term between frequency and prior frequency can help determine whether the elasticity differs between previously frequent and infrequent routes.

*6.3. Sensitivity to Prior-Frequency*

Table 7 shows the fixed-effects model results where frequency is interacted with prior frequency. The $\log(\text{Pop}_{it} + \text{Job}_{it})$ and linear time coefficients are almost identical to the previous model. The elasticity is the sum of its two first terms, $\beta_1 + \beta_2 \log(\text{Freq}_{it_0})$. The estimate of $\beta_1$ is far off from the estimate in Table 6 because the term is not centered. Its interpretation should assume that $\log(\text{Freq}_{it_0}) = 0$, i.e. that $\text{Freq}_{it_0} = 1$.

Figure 5 shows the elasticity of ridership to frequency as a function of prior frequency. Confidence bands of one standard deviation surround the estimated elasticity. Since the elasticity term includes two components, one fixed, $\beta_1$, and one based on frequency in the first markup, $\beta_2 \log(\text{Freq}_{it_0})$, the combined standard deviation was obtained using the Delta Method (Oehlert, 1992). In Portland, Miami, and Atlanta, elasticity is greatest on low-frequency routes, while in Minneapolis/St-Paul elasticity is greatest on high-frequency routes. These results indicate that in all agencies besides Minneapolis, each percentage increase in frequency will produce a greater percentage increase in ridership on routes that were previously infrequent than on routes that already had high-frequency. The reason why Minneapolis deviates from its peers is possibly due to the looser connection between frequency and productivity at one point in time exhibited in Figure 3. It is likely that Metro Transit has historically invested in service coverage over concentration, thereby maintaining some potential growth on the most frequent routes, which has been exhausted in other agencies.

Table 7: Fixed Effects Model With Prior Frequency

|  | Response Variable: | $\text{Rid}_{it}$ | | |
| --- | --- | --- | --- | --- |
|  | Portland | Miami | Minneapolis/St-Paul | Atlanta |
| $\log(\text{Freq}_{it})$ | 1.16 (0.12)*** | 1.66 (0.24)*** | 0.05 (0.15) | 1.50 (0.21)*** |
| $\log(\text{Freq}_{it_0}) * \log(\text{Freq}_{it})$ | −0.12 (0.03)*** | −0.23 (0.06)*** | 0.19 (0.04)*** | −0.22 (0.05)*** |
| $\log(\text{Pop}_{it} + \text{Job}_{it})$ | −0.02 (0.02) | 0.03 (0.02) | 0.04 (0.02) | 0.11 (0.03)*** |
| t | −0.02 (0.00)*** | −0.05 (0.00)*** | −0.03 (0.00)*** | −0.07 (0.00)*** |
| Log-Likelihood | −35162.98 | −15134.52 | −34290.85 | −11412.27 |
| Num. obs. | 17640 | 10575 | 18256 | 7283 |
| n | 874 | 1165 | 959 | 718 |
| T | 21 | 10 | 21 | 12 |

***$p < 0.001$; **$p < 0.01$; *$p < 0.05$; ·$p < 0.1$

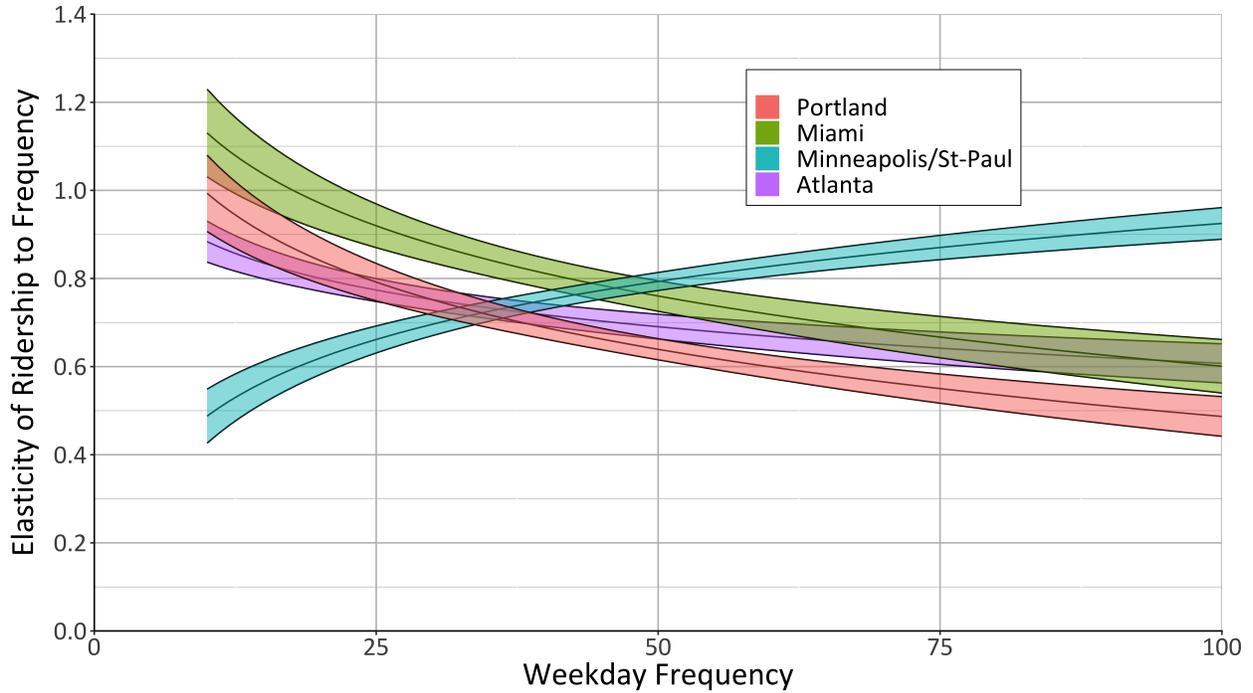

Figure 5: Elasticity of ridership to frequency as a function of prior frequency

## 7. Conclusion and Future Research

This study reveals the relationship between bus ridership and frequency on a highly disaggregated level. In a cross-sectional analysis, we found that the most frequent routes at one point-in-time are also the most productive (passengers per vehicle-trip) in all four transit agencies. This relationship, however, contains both the exogenous effect of ridership on frequency, which we are interested in, and the reverse causality. Through decades of service planning, the bus network reflects the spatial distribution of travel demand. To control for time-invariant endogeneity and other unobserved heterogeneity **between** route-segments, we developed a Poisson fixed-effects model, which captures the variation in ridership **within** each route-segment over time. The results indicate that the weekday ridership elasticity to frequency ranges from 0.66 to 0.78 for all agencies. We also find that in Portland, Miami, and Atlanta, the most frequent routes are the least elastic while the opposite is true in Minneapolis/St-Paul. Overall, three broad conclusions can be drawn from the analysis presented herein:

- The wide difference in estimated elasticity between the cross-section and fixed-effects models, which accounts for the endogeneity and unobserved heterogeneity, indicates that **the demand for bus service is more concentrated than the supply**. As shown in the cross-sectional analysis of Subsection 6.1, frequent routes serve more passengers per trip than lower-frequency routes.

- For all agencies studied, **ridership is inelastic to frequency**, as shown in Subsection 6.2.

Therefore, each marginal trip added to a route will typically generate less ridership than the average vehicle already on the route. In other words, the expected ridership returns from frequency are diminishing.

- If transit agencies were to increase frequency on all routes by the same proportion, the productivity advantage of frequent routes would decline in Portland, Miami, and Atlanta and increase in Minneapolis/St-Paul as shown in Subsection 6.3.

The main limitation of this study is that we do not account for potential time-varying endogeneity. If transit planners readjusted service in response to the organic change in travel demand between 2012 and 2018 that happened independently of service frequency, population, and jobs, then elasticity estimates may be inflated in the fixed-effects models (Wooldridge, 2002, §9). We expect, however, this time-varying endogeneity to be relatively small. Even when planners do allocate service in response to changes in demand, the potential time-varying endogeneity does not undermine the main conclusions of this paper. If anything, inflated elasticity estimates caused by time-varying endogeneity would bolster our conclusion that ridership is inelastic to frequency.

Service allocation is by far the most important lever transit agencies have available to affect ridership without changes in operating budgets. The results in this paper can be used to anticipate the expected ridership change from cutting service on one route to prioritize another solely based on prior frequency and ridership of both routes, as shown in Section 5.3. Specifically, the estimated marginal ridership of each additional vehicle-trip can be obtained by simply multiplying a route's current productivity by its elasticity. Therefore, the insights unraveled in this study can serve as inputs for travel demand models evaluating broader changes to the bus network. The results can also support decisions to prioritize service coverage or concentration on a case by case basis to attain the best possible compromise.

The service allocation problem consists in setting frequencies throughout the bus network with the objective of minimizing a combination of waiting costs for passengers and operating costs for the agency (Mohring, 1972). The method was extended by Furth and Wilson (1981) to consider the societal benefit of transit ridership in the objective function based on elasticities from Mayworm et al. (1980). More recent research has used ridership elasticity to frequency as an input parameter (Verbas and Mahmassani, 2013). The optimal service allocation policy was found to be sensitive to the assumed elasticity. The elasticities modeled in this paper can support these types of service allocation optimization tools. Future research could take the relationship between elasticity and prior frequency into account to attain even greater productivity.

The methodology applied to this problem for the first time can serve as a framework to support future research in transit demand using observational data. While frequency is not the only factor affecting ridership, it determines the feasibility, travel time, and reliability of transit trips. Several studies had investigated how the change in service levels could explain changes in ridership over time at the metropolitan area or transit agency level (Taylor et al., 2009; Chen et al., 2011; Boisjoly et al., 2018; Watkins et al., 2019); others had explored the correlation between frequency and ridership on a hyper-local level as a cross-section (Dill et al., 2013; Chakour and Eluru, 2016; Aston et al., 2020b). For the first time, a Poisson Fixed-Effects model revealed the association between the change in service frequency and bus ridership over time on a hyper-local level.

The model presented in this paper could be extended in future research to consider the ridership impact of nearby service changes, from slight bus schedule adjustment to heavy rail station opening. Future research should also explore the sensitivity of demand to service quantity by time-of-day and day of the week. Knowing in which time-period ridership has declined the most could help understand the underlying causes. Knowing when ridership is most elastic to frequency could lead to service allocation strategies that can maximize ridership.

Several research questions remain unaddressed. In particular, service changes do not explain the current nation-wide ridership crisis. The competition from other modes including dynamic mobility companies may also affect the demand for buses on a local level. They may ultimately explain the ridership change at the regional and national levels. But their effects are necessarily lower order as they tend to drift slowly over time. Therefore, the elasticity to frequency is necessary to understand the underlying causes of ridership change. The approach presented in this paper will allow future research to understand the causes of ridership decline and identify strategies to reverse the trend.

## 8. Aknowledgments

The authors wish to extend heartfelt gratitude for staff members at TriMet, Miami-Dade Transit, Metro Transit, and MARTA for providing us with the data and helping us understand their transit systems. In particular, we thank Nathan Banks at Trimet, Eric Lind, Janet Hopper, and John Levin at Metro Transit, Esther Frometa-Spring at Miami-Dade Transit, and Charlotte Owusu-Smart, Robin Salter, and Robert Goodwin at MARTA. We also thank Thomas Kowalsky from Urban Transportation Associates for his help navigating the data and understanding the technology. Jorge Laval provided valuable feedback on the modeling decisions. Finally, we are particularly grateful for the help of Taylor Gibbs in the data preparation stage of this research.

## Appendix A. Missing Trip Sensitivity Analysis

The results of Section 6 are revisited in this Appendix based on the number of missing trips from GTFS in APC. Figure A.6 shows (a) the sample size and (b) the estimated ridership elasticity to frequency based on the maximum number of missing trips. In all four agencies, allowing segments with missing trips in the analysis marginally increases the same size and causes slight fluctuations in the elasticity coefficient estimate.

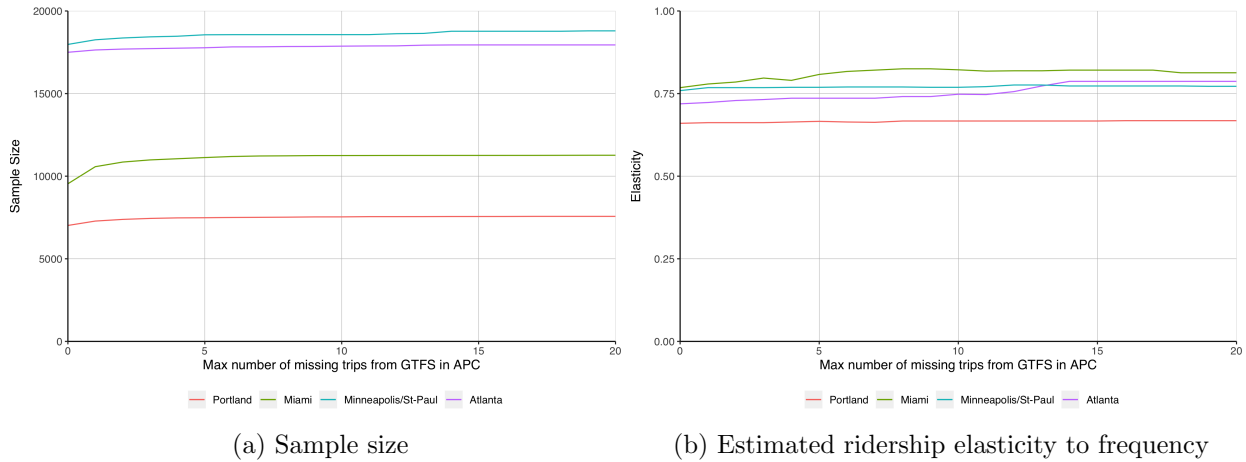

(a) Sample size  (b) Estimated ridership elasticity to frequency

Figure A.6: Sensitivity analysis to maximum allowable number of missing trips